\documentclass{article}

\usepackage{graphicx}
\usepackage{amsmath}

\title{Transient fluctuation of the prosperity of firms in a network economy}

\author{Yoshiharu Maeno \\ Social Design Group \\ email: maeno.yoshiharu@socialdesigngroup.com}

\begin{document}

\maketitle

\begin{abstract}
The transient fluctuation of the prosperity of firms in a network economy is investigated with an abstract stochastic model. The model describes the profit which firms make when they sell materials to a firm which produces a product and the fixed cost expense to the firms to produce those materials and product. The formulae for this model are parallel to those for population dynamics. The swinging changes in the fluctuation in the transient state from the initial growth to the final steady state are the consequence of a topology-dependent time trial competition between the profitable interactions and expense. The firm in a sparse random network economy is more likely to go bankrupt than expected from the value of the limit of the fluctuation in the steady state, and there is a risk of failing to reach by far the less fluctuating steady state.
\end{abstract}

\section{Introduction}

Firms are put under the selection pressure in a network economy. Every firm searches for the most profitable pattern of interactions with other firms for the trades of goods and credits to survive. Financial distress, however, transmits along the interactions as well. As a firm declines in sales for whatsoever reasons, bad debts of the firms which sell goods to that firm tend to increase. The firms have a hard time to manage their financing, also decline, and some of them even end in bankruptcy. The bad debts increase still more. The risk that financial distress emerges in a network economy and transmits widely is called systemic risk in macro-economics. The pattern of interactions has a decisive impact on both the financial distress and the bankruptcy of firms.

A supply chain and a business cluster are often found highly efficient sub-systems in a real network economy. A supply chain is a group of firms in a diverse range of industry sectors. For example, a firm in the finished goods sector purchases components from a firm in the component sector, and the firm purchases materials from a firm in the materials sector in turn. As goods are moved downstream, money is transferred upstream in the opposite direction. The topological pattern of these interactions along a stream is a linear chain most simply. A business cluster is a geographic concentration of firms in closely related industry sectors. Empirically, firms develop reciprocally profitable interactions with each other, raise each otherfs prosperity collectively, and increase the productivity as a whole. Win-win games ensue there. The topological pattern of such reciprocal interactions is a ring most simply. Do chain and ring topologies have any special properties to keep financial distress from emerging and transmitting? How does the prosperity of individual firms evolve temporarily under such patterns of interactions?

In this study, the transient fluctuation of the prosperity of firms is investigated with an abstract stochastic model. The model describes the profit which firms make when they sell materials to a firm which produces a product, given a fixed pattern of the network for the trades of goods, and the fixed cost expense to the firms to produce those materials and product. The mathematical formulae of the model are parallel to those for population dynamics. The prosperity and the risk of bankruptcy of a firm resemble the population and the risk of extinction of a species. The analytic solution of the fluctuation is presented for two building blocks of the entire pattern of interactions, namely, chain and ring with homogeneous profitability and expense parameters. The fluctuation and its temporal correlation are analyzed numerically for more general random pattern of interactions with heterogeneous profitability and expense parameters.

\section{Related works}

The contagion network model of three industry sectors describes the growth of firms \cite{Del10}. The model combines the change in the amount of trades from firms in the materials industry sector to firms in the finished goods sector, and the change in the interest rate of loans from banks in the finance sector to the firms in both industry sectors. The systemic risk emerges from the cycle that the cumulative loss of firms causes a sharp rise in the interest rate for the firms to raise funds from banks, and increases the loss still more. When the net worth of a firm reaches zero by netting the cumulative loss, the firm goes bankrupt. In the model, the bankrupt firm is replaced with a new firm. The firms and banks may change selectively the topology of the network for the trades of goods and credits as the interest rate rises and falls \cite{Del06}. The price of the finished goods is a stochastic exogenous variable. A large decline of price represents the shock which is exerted by a macro-economic environment. The steady state distribution of the net worth and growth rate of firms is obtained with simulations. These studies analyze how the avalanche of bankruptcies of firms ensues from exogenous macro-economic shocks, namely, a sudden fall of demands and a resulting decline of the price.

Similar contagion network models are applied to analyze the transmission of distress among banks in an inter-bank credit network \cite{Hal11}, \cite{Gai10}, \cite{Mar10}, \cite{Bey07}. These studies are of particular interest worldwide since the recent financial crises. They reveal how the avalanche of bankruptcies of banks ensues from exogenous macro-economic shocks like a sudden decrease of funds transfer requests from client firms, the collapse of an asset market, and the disappearing liquidity of the assets in which banks make an investment. None of these studies, however, addresses the problem on how the probability of the bankruptcy of a firm or a bank, which ignites the avalanche of bankruptcies, is correlated with endogenous factors like the topology of a network and the fluctuating trades of goods or credits between them.

In terms of mathematical formulae, the interactions between firms are parallel to the interactions between species in the population dynamics \cite{All11}, \cite{Lar11}, \cite{Wu11}, \cite{Meh09}. The bankruptcy of a firm when the prosperity reaches zero resembles the extinction of a species when the population reaches zero. Jain and Krishna's linear catalytic network is a weighted digraph whose vertex and arc represent a species and the presence of catalysis from one species to another \cite{Jai01}, \cite{Jai98}. The weight of an individual arc is the strength of the catalysis. The steady state distribution of the relative populations of species is a function of the topology of the digraph. Species of the smallest population in the steady state are dismissed and replaced by new species. This results in an update of the topology of the digraph. An auto-catalytic set appears by chance after an update \cite{Jai02a}, \cite{Jai02b}. It is a set of catalyzed species which governs the proliferation, extinction, and recovery of the entire system of species. Many real systems, however, do not reach the steady state quickly, but remain in a transient state. The strength of the selection pressure on individual species and the resulting probability of their extinction are not predicted from their relative populations in the steady state which these studies reveal.

\section{Stochastic model}

A stochastic model for a network economy is presented in this section. A fixed digraph represents the entire network economy. The vertex $v_{i}$ is the $i$-th firm for $i=0,1,\cdots,N-1$. $N$ is the number of firms. The arc $v_{j} \rightarrow v_{i}$ represents the interaction from the vertex $v_{j}$ to $v_{i}$. An arc $v_{i} \rightarrow v_{i}$ is a loop.

Continuous variables $y_{i}(t)$ is the prosperity of the $i$-th firm at continuous time $t$. The prosperity is an abstract quantity to represent the net worth, or the cumulative profit, of the firm. This is parallel to the population of a species in population dynamics. Assume two types of abstract interactions between firms. As the $j$-th firm produces products, its purchase of materials gives rise to the increase of the prosperities of a set of firms. The rate of change of $y_{i}$ is given by $c_{ij} y_{j}(t)$ where the profitability parameters $c_{ij}$ are constant. This resembles the $j$-th species (prey) which catalyzes the proliferation of the $i$-th species (predators). The non-zero elements of the matrix $c_{ij}$ represent the arcs $\mbox{\boldmath{$E$}}=\{v_{j} \rightarrow v_{i}\}$. The decrease of $y_{i}$ results from the fixed cost expense by paying for the labor to produce products. The rate is given by $-d_{i} y_{i}(t)$ where the expense parameters $d_{i}$ are constant. This resembles the individual deaths of the $i$-th species in the population dynamics.

Stochasticity ensues from an unpredictably irregular temporal pattern of these interactions. If the probability of an interaction per unit time is constant, the number of interactions in a given time interval obeys a Poisson distribution \cite{Kom11}, where the amplitude of deviation is equal to the mean. The time evolution of $y_{i}(t)$ is given by a system of Langevin equations in eq.(\ref{langevin}). The functional forms of the Gaussian white noises, ${}^{{\rm c}}\xi_{j}(t)$ and ${}^{{\rm d}}\xi_{i}(t)$, are not known.
\begin{equation}
\frac{{\rm d}y_{i}(t)}{{\rm d}t} = \sum_{j=0}^{N-1} (c_{ij} y_{j}(t) + \sqrt{c_{ij} y_{j}(t)} {}^{{\rm c}}\xi_{j}(t)) - d_{i} y_{i}(t) - \sqrt{d_{i} y_{i}(t)} {}^{{\rm d}}\xi_{i}(t).
\label{langevin}
\end{equation}

The solutions of eq.(\ref{langevin}) are described equivalently by the joint probability density function $P(\mbox{\boldmath{$y$}},t)$ of a random variable $\mbox{\boldmath{$y$}} = (y_{0},\cdots,y_{N-1})$ at $t$. Its time evolution is given by a Fokker-Planck equation \cite{Fa11}, \cite{Dob12} in eq.(\ref{FokkerPlanck}).
\begin{equation}
\frac{\partial P(\mbox{\boldmath{$y$}},t)}{\partial t} = -\sum_{i} \frac{\partial}{\partial y_{i}} A_{i} P(\mbox{\boldmath{$y$}},t) + \frac{1}{2} \sum_{i,j} \frac{\partial^{2}}{\partial y_{i} \partial y_{j}} B_{ij}  P(\mbox{\boldmath{$y$}},t).
\label{FokkerPlanck}
\end{equation}

The elements of the drift vector $\mbox{\boldmath{$A$}}$ and the diffusion matrix $\mbox{\boldmath{$B$}}$ is given by eq.(\ref{FPequationA}) and (\ref{FPequationB}). 
\begin{equation}
A_{i} = \sum_{j=0}^{N-1} \tilde{A}_{ij} y_{j} = \sum_{j=0}^{N-1} (c_{ij}-d_{i}\delta_{ij})y_{j}.
\label{FPequationA}
\end{equation}
\begin{equation}
B_{ij} = \sum_{k=0}^{N-1} \{(c_{ik}+d_{i}\delta_{ik})\delta_{ij}+\sqrt{c_{ik}c_{jk}}(1-\delta_{ij})\} y_{k}.
\label{FPequationB}
\end{equation}

\section{Analytic solution}

A comparative analysis of the impact of topologies on the fluctuation of prosperity is presented in this section. The digraph for the entire network economy includes a number of chain and ring sub-digraphs. A ring sub-graph is the simplest abstract form of collectively reciprocal interactions which are seen in a business cluster. This is an auto-catalytic set in the population dynamics. A chain sub-graph represents an abstract supply-chain where reciprocal interactions do not work. The nature of these building blocks is the basis to understand the impact of the topology of general random digraphs. The formulae for the fluctuation are derived for these two analytically tractable sub-digraphs with homogeneous profitability and expense parameters.

The arcs of the chain are $\mbox{\boldmath{$E$}}_{{\rm C}}=\{v_{0} \rightarrow v_{1}, v_{1} \rightarrow v_{2}, \cdots, v_{N-2} \rightarrow v_{N-1} \}$. The vertex $v_{0}$ is the most downstream purchaser, and $v_{N-1}$ the most upstream seller. Assume homogeneous parameters, $c_{ij}=c \geq 0$ for $j=i$, $c_{ij}=c' > 0$ for $j=i-1$, $c_{ij}=0$ for the other values of $j$, and $d_{i}=d \geq 0$ for all $i$. Self-sustaining profitability, a firm with reciprocal subsidiaries or stably profit-making assets, is meant when $c>0$ (loops $\{v_{0} \rightarrow v_{0}, \cdots, v_{N-1} \rightarrow v_{N-1}\}$). This is self-catalysis in the population dynamics. The arcs of the ring are $\mbox{\boldmath{$E$}}_{{\rm R}}=\mbox{\boldmath{$E$}}_{{\rm C}} \cup \{ v_{N-1} \rightarrow v_{0} \}$. Assume $c_{ij}=c$ for $j=i$, $c_{ij}=c'$ for $j=i-1 \% N$, $c_{ij}=0$ for the others. The modulo operation $i-1 \% N$ outputs the remainder on dividing $i-1$ by $N$.

The time evolution of the moments of $\mbox{\boldmath{$y$}}$ is derived from the Fokker-Planck equation \cite{Mae10}, \cite{Mae11}. The first order moments, ${}^{1}\mu_{i}(t) = \langle y_{i} \rangle_{t}$, is the ensemble average of $y_{i}$ at $t$. The initial condition is ${}^{1}\mu_{i}(0) = y_{i}(0)$. It is given by a function of $t$ in eq.(\ref{mu1formula}). It increases monotonically if $c \geq d$.
\begin{equation}
{}^{1}\mu_{i}(t) = [ \exp (\tilde{\mbox{\boldmath{$A$}}} t) \mbox{\boldmath{$y$}}(0) ]_{i}
= \sum_{Q(i)} \frac{y_{q}(0) c'^{q'}}{q'!} t^{q'} e^{(c-d)t}.
\label{mu1formula}
\end{equation}

The bounds of summation with non-negative integer index variables $q$ and $q'$ are given by the set $Q(i)$ in Table \ref{Qtable}. For the ring, $q'$ is given by a Diophantine equation $q' \equiv i-q\ {\rm mod}\ N$ whose solution is $q'=i-q, i-q+N, \cdots$ if $i \geq q$, and $q'=i-q+N, i-q+2N, \cdots$ if $i<q$. Significant profitability from the nearest neighbor firm is ignited when $c't \sim 1$. The ignition time is $t_{{\rm p}} \sim 1/c'$. The terms of $q' \geq N$ represent reciprocal profitability. They are ignited when $(c't)^{N}/N! \sim 1$. The ignition time is $t_{{\rm r}}(N) \sim N/ec'$ if $N \gg 1$.

The second order moments about the mean (covariance), ${}^{2}\mu_{ij}(t) = \langle y_{i} y_{j} \rangle_{t} - {}^{1}\mu_{i}(t) {}^{1}\mu_{j}(t)$, is given by eq.(\ref{mu2formula}). The initial condition is ${}^{2}\mu_{ij}(t) = 0$. The operand functions of the three summation operators $\sum$ are the same.
\begin{eqnarray}
{}^{2}\mu_{ij}(t) &=& \int_{0}^{t} (\exp \tilde{\mbox{\boldmath{$A$}}} (t-t') \langle \mbox{\boldmath{$B$}} \rangle_{t'} \exp \tilde{\mbox{\boldmath{$A$}}}^{{\rm T}} (t-t') )_{ij} {\rm d}t' \nonumber \\
&=& \{ (c+d) \sum_{Q_{1}(i,j)} + c' \sum_{Q_{2}(i,j)} + \sqrt{cc'} \sum_{Q_{3}(i,j) \cup Q_{4}(i,j)} \} \nonumber \\
& & \frac{y_{q}(0) c'^{q'+q''+q'''} }{q'! q''! q'''!} T_{q'',q'+q'''}(t).
\label{mu2formula}
\end{eqnarray}

\begin{table}
\caption{Set $Q(i)$ which gives the bounds of summation with $q$ and $q'$ for ${}^{1}\mu_{i}(t)$ in eq.(\ref{mu1formula}).}
\begin{center}
\begin{tabular}{|c|c|c|}
\hline
 & chain & ring \\
\hline
$Q(i)$ & $\begin{matrix} 0 \leq q \leq i \\ q'=i-q \end{matrix}$ & $\begin{matrix} 0 \leq q \leq N-1 \\ q' \equiv i-q\ {\rm mod}\ N\end{matrix}$ \\
\hline
\end{tabular}
\end{center}
\label{Qtable}
\end{table}

\begin{table}
\caption{Sets $Q_{1}(i,j)$ through $Q_{4}(i,j)$ which give the bounds of the $\sum$ operators with $q$, $q'$, $q''$, and $q'''$ for ${}^{2}\mu_{ij}(t)$ in eq.(\ref{mu2formula}).}
\begin{center}
\begin{tabular}{|c|c|c|}
\hline
 & chain & ring \\
\hline
$Q_{1}(i,j)$ & $\begin{matrix} 0 \leq k \leq i,j \\ 0 \leq q \leq k \\ q'=i-k \\ q''=k-q \\ q'''=j-k \end{matrix}$ & $\begin{matrix} 0 \leq k \leq N-1 \\ 0 \leq q \leq N-1 \\ q' \equiv i-k\ {\rm mod}\ N \\ q'' \equiv k-q\ {\rm mod}\ N \\ q''' \equiv j-k\ {\rm mod}\ N \end{matrix}$ \\
\hline
$Q_{2}(i,j)$ & $\begin{matrix} 1 \leq k \leq i,j \\ 0 \leq q \leq k-1 \\ q'=i-k \\ q''=k-q-1 \\ q'''=j-k \end{matrix}$ & $\begin{matrix} 0 \leq k \leq N-1 \\ 0 \leq q \leq N-1 \\ q' \equiv i-k\ {\rm mod}\ N \\ q'' \equiv k-q-1\ {\rm mod}\ N \\ q''' \equiv j-k\ {\rm mod}\ N \end{matrix}$ \\
\hline
$Q_{3}(i,j)$ & $\begin{matrix} 1 \leq k \leq i,j+1 \\ 0 \leq q \leq k-1 \\ q'=i-k \\ q''=k-q-1 \\ q'''=j-k+1  \end{matrix}$ & $\begin{matrix} 0 \leq k \leq N-1 \\ 0 \leq q \leq N-1 \\ q' \equiv i-k\ {\rm mod}\ N \\ q'' \equiv k-q-1\ {\rm mod}\ N \\ q''' \equiv j-k+1\ {\rm mod}\ N \end{matrix}$ \\
\hline
$Q_{4}(i,j)$ & $\begin{matrix} 0 \leq k \leq i,j-1 \\ 0 \leq q \leq k \\ q'=i-k \\ q''=k-q \\ q'''=j-k-1 \end{matrix}$ & $\begin{matrix} 0 \leq k \leq N-1 \\ 0 \leq q \leq N-1 \\ q' \equiv i-k\ {\rm mod}\ N \\ q'' \equiv k-q\ {\rm mod}\ N \\ q''' \equiv j-k-1\ {\rm mod}\ N \end{matrix}$ \\
\hline
\end{tabular}
\end{center}
\label{Q1234table}
\end{table}

The bounds of the $\sum$ operators with non-negative integer index variables $q$, $q'$, $q''$, and $q'''$ are given by the sets $Q_{1}(i,j)$ through $Q_{4}(i,j)$ in Table \ref{Q1234table}. $T_{r,r'}(t)$ is specified by two non-negative integers $r$ and $r'$, and given by eq.(\ref{Trr'}).
\begin{eqnarray}
T_{r,r'}(t) &=& \int_{0}^{t} t'^{r} (t-t')^{r'} e^{(c-d)(2t-t')} {\rm d}t' \nonumber \\
&=& \sum_{r''=0}^{r'} \frac{ _{r'}\mathrm{C}_{r''} (-1)^{r'-r''} (r+r'-r'')! }{(c-d)^{r+r'-r''+1}} t^{r''} \nonumber \\
&\times& \{e^{2(c-d)t} - \sum_{r'''=0}^{r+r'-r''} \frac{(c-d)^{r'''} t^{r'''}}{r'''!} e^{(c-d)t} \}.
\label{Trr'}
\end{eqnarray}

Eq.(\ref{mu1formula}) and (\ref{mu2formula}) hold true for both the chain and ring sub-graphs. The effect of the reciprocal profitability of a ring appears in the bounds of summation in Table \ref{Qtable} and \ref{Q1234table}. The 3rd order moments, ${}^{3}\mu_{ijk}(t)$, and higher order moments are derived similarly. The explicit formula for $P(\mbox{\boldmath{$y$}},t)$ is obtained by a multi-variate Edgeworth series \cite{Bal88}. The fluctuation is defined by eq.(\ref{fluc}). This quantity is also used in quantifying the noise in a changing environment \cite{Hil11}, and the variability of spreading phenomena \cite{Cre06}.
\begin{equation}
F_{ij}(t) = \sqrt{ \frac{{}^{2}\mu_{ij}(t) }{ {}^{1}\mu_{i}(t) {}^{1}\mu_{j}(t)}  }.
\label{fluc}
\end{equation}

\begin{figure}
\begin{center}
\includegraphics[scale=0.65,angle=0]{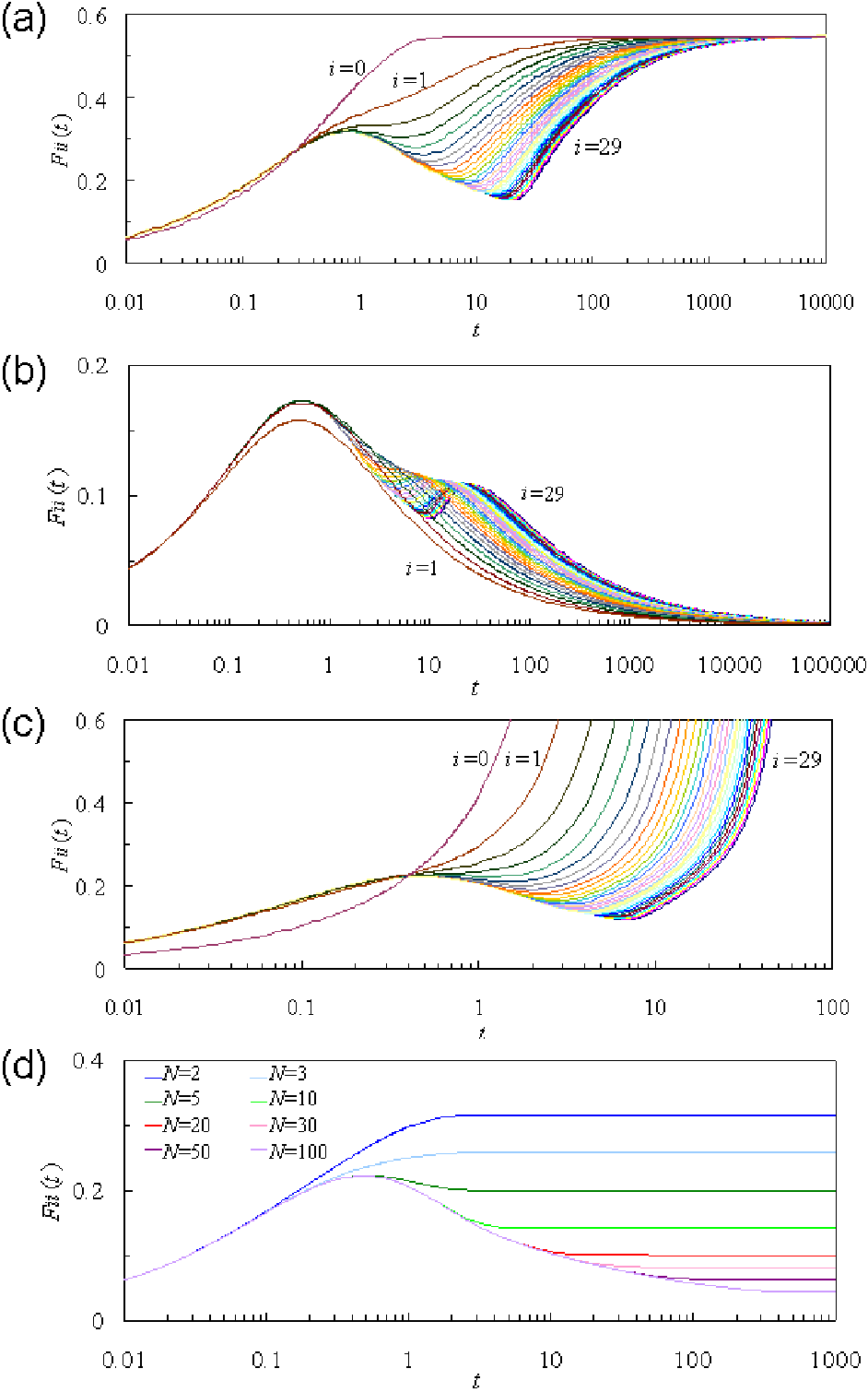}
\end{center}
\caption{$F_{ii}(t)$ as a function of $t$. (a) chain of $N=30$ ($i$=0 to 29) when $c=2$ (self-sustaining profitability), $c'=1$, and $d=1$. For all $i$, $y_{i}(0)=10$. (b) chain when $c=0$, $c'=2$, and $d=0$ (no expense). Nothing happens for $i=0$. (c) chain when $c=0$, $c'=3$, and $d=1$. (d) rings of $N=2$, 3, 5, 10, 20, 30, 50, and 100 when $c=0$, $c'=3$, and $d=1$.}
\label{pre301s}
\end{figure}

Figure \ref{pre301s} shows $F_{ii}(t)$ for ring and chain sub-graphs as a function of $t$. For a ring, $c'>0$ alone does not contain the diverging fluctuation. $F_{ii}$ converges if $c>0$ or $d=0$. Downstream vertices ($v_{0}$ etc.) suffer from larger fluctuation then upstream vertices ($v_{N-1}$ etc.) if $d>0$. For a ring, the fluctuation in the transient state is larger than that in the steady state when $N \geq 5$.

The limit for the chain is given by eq.(\ref{openfluc}). It does not depend on any of $i$, $j$, $N$, $c'$. For any $i$, $j$, the dependence on the initial condition is $y_{0}(0)^{-1/2}$. Without self-sustaining profitability, the limit diverges unless the expense for production is absent. Convergence is slow ($\sim t^{-1}$), and particularly slow ($\sim t^{-\frac{1}{2}}$) when $c=d=0$. Eq.(\ref{openfluc}) holds true not only for the linear chain discussed so far, but for a branching chain. The branching chain is a tree sub-graph where any vertices neighbor one or more upstream vertices and a downstream vertex. A tree is such a supply chain in an assembly industry as an electronics company purchasing from circuit module vendors, each of which purchases from semiconductor chip makers. The chain and tree sub-graphs look alike because the growth of prosperity is not influenced by upstream vertices.
\begin{equation}
\lim_{t \rightarrow \infty} F_{ij} = 
\begin{cases}
\sqrt{ \frac{1}{y_{0}(0)} \frac{c+d}{c-d} } & c-d > 0 \\
\infty & c-d \leq 0\ {\rm except}\ c=d=0 \\
0 & c=d=0
\end{cases}.
\label{openfluc}
\end{equation}

The limit for the ring is given by eq.(\ref{closefluc}). It does not depend on $i$, $j$. For any $i$, $j$, the dependence on the initial condition is $(\sum_{q} y_{q}(0))^{-1/2}$. Because the limit in eq.(\ref{closefluc}) tends to decrease as $N$ increases ($N^{-1/2}$), the gap between the limits for the chain and ring is more conspicuous as $N$ increases. As $c$, $c'$ decreases or $d$ increases, $F$ increases. Even without self-sustaining profitability, the limit is finite as far as $d < c'$ is satisfied.
\begin{equation}
\lim_{t \rightarrow \infty} F_{ij} = 
\begin{cases}
\sqrt{ \frac{1}{\sum_{q} y_{q}(0)}\frac{c+c'+d+2\sqrt{cc'}}{c+c'-d} } & c-d > -c' \\
\infty & c-d \leq -c'
\end{cases}.
\label{closefluc}
\end{equation}

At small $t$, $F_{ij}$ is given by eq.(\ref{fluc_t0}). The time evolution of $F_{ij}$ for chains is equal to that of rings of any length with the same parameters. Vertices are exposed to nearly identical trends of growing fluctuation. None of these sub-digraphs are discriminated from each other in the digraph with homogeneous parameters in the early transient state.
\begin{equation}
F_{ij} \approx \sqrt{t} \ [ \frac{ \{ (c+d)y_{i}(0) + c' y_{i-1}(0) \} \delta_{ij}}{ y_{i}(0) y_{j}(0) } + \frac{ \sqrt{cc'} (y_{i}(0) \delta_{i+1\ j} + y_{i-1}(0) \delta_{i-1\ j}) }{ y_{i}(0) y_{j}(0) } ]^{-\frac{1}{2}}.
\label{fluc_t0}
\end{equation}

\section{Numerical analysis}

\begin{figure}
\begin{center}
\includegraphics[scale=0.65,angle=0]{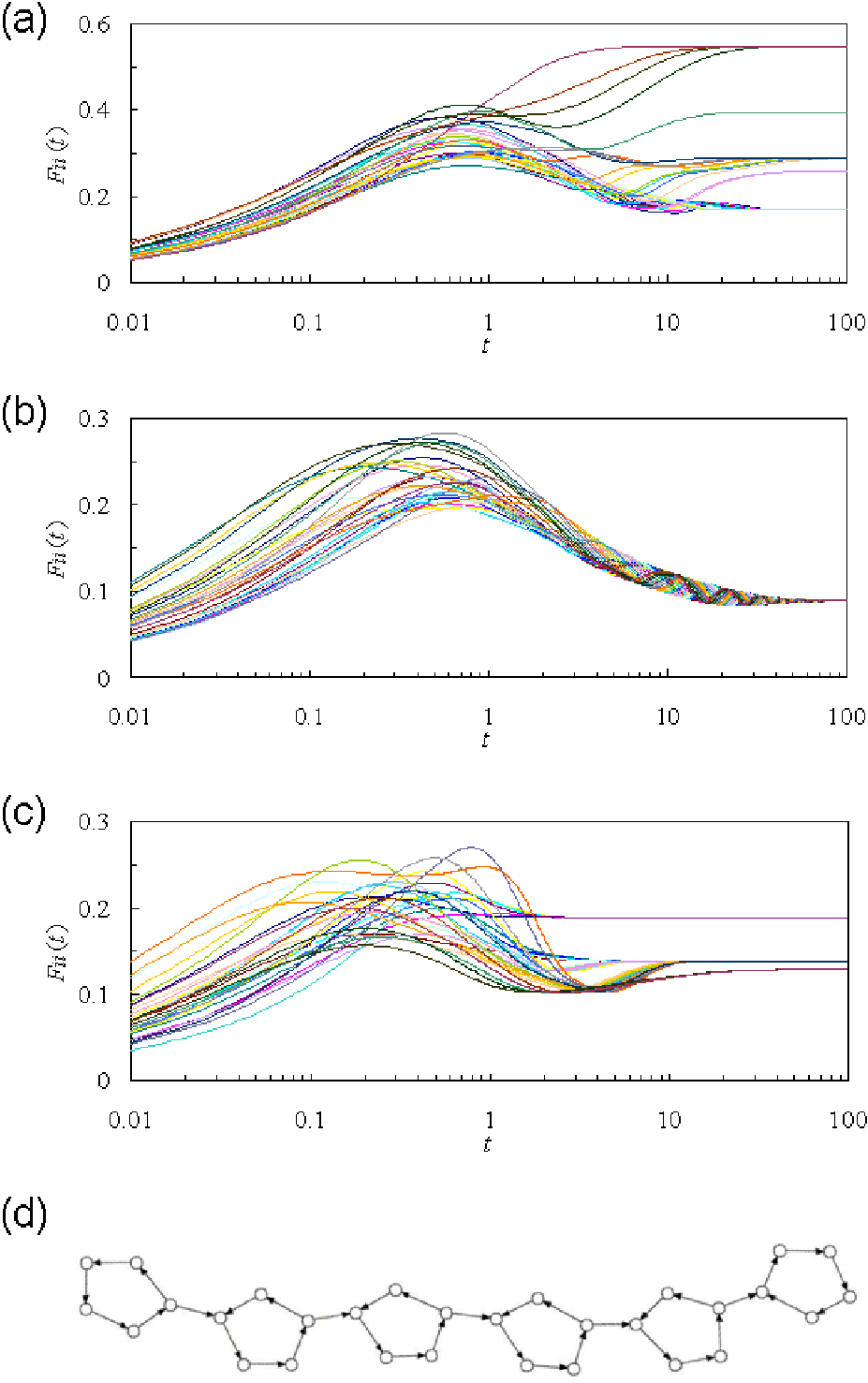}
\end{center}
\caption{$F_{ii}(t)$ for $i=0$ to 29 ($N=30$) as a function of $t$ when $c_{ij}$, $d_{i}$, and $y_{i}(0)$ are chosen randomly. (a) Chain with $\hat{c}=2$ for $j=i$ (self-sustaining profitability), 1 for $j=i-1$, and 0 for the other $j$. (b) Ring with $\hat{c}=3$ for $j=i-1\%N$ and 0 for the other $j$. (c) Chain of rings whose topology is shown in (d) with $\hat{c}=3$ if an arc is present and 0 otherwise. For the all graphs, $\hat{d}=1$, $\hat{y}(0)=10$, and $h=0.5$.}
\label{pre302s}
\end{figure}

The nature of more general random digraphs is presented in this section. First, the fluctuation for digraphs with heterogeneous profitability and expense parameters is analyzed by evaluating the integral in eq.(\ref{mu2formula}) numerically. Fig. \ref{pre302s} shows $F_{ii}(t)$ for a chain, a ring, and a chain of rings when $c_{ij}$ is chosen randomly between $(1-h)\hat{c}$ and $(1+h)\hat{c}$, given the median $\hat{c}$ and heterogeneity parameter $h$. Given $\hat{d}$ and $\hat{y}(0)$, $d_{i}$ and $y_{i}(0)$ are chosen randomly in the same manner.

In (a), most curves for $F$ take a local maximum at about $t=t_{{\rm p}}$ which is followed by a few local extrema. The number of the extrema depends on vertices. The curve for an downstream vertex (with $i$) along the chain tends to attract those for its upstream neighbor vertex (with $i+1$ etc.). $F_{ll} \leq F_{kk}$ holds true for any $l > k$ at large $t$. The degenerate limit in eq.(\ref{openfluc}) is split into multiple values. The number of the limits depends on the topology and the value of the parameters. Similar numerical examples show if the chains without self-sustaining profitability are shorter, the expense starts surpassing the profitability and an infinitely growing trend of $F$ appears earlier. In (b), the curves for $F$ for the ring oscillate regularly after about $t=t_{{\rm r}}$. Crests and troughs of the profitability arise locally, and transmit along the ring as a travelling wave. This profit-making wave does not impair the convergence to the single limit. Generally, if the rings are shorter, the reciprocal profitability is ignited and $F$ starts converging earlier. In (c), the fluctuation resembles that for the chain, but the number of limits tends to be smaller. The swinging changes in $F$ at about $t=t_{{\rm p}}$ until about $t=t_{{\rm r}}$ represent the transient state from the common initial growth to the final steady state.

Next, the dependence of the transient state on the number of arcs and rewiring of random digraphs, and on the heterogeneity of profitability and expense parameters is investigated with the temporal correlation of $F_{ii}(t)$. The correlation is the Pearson's product-moment correlation coefficient of $F$ between the inception of growth at $t_{1}=0.01$ and $t$. It is defined by eq.(\ref{corr}). $\bar{F}(t)$ is the average of $F_{ii}$ over the all vertices at $t$.
\begin{equation}
R(t) = \frac{\sum_{i} (F_{ii}(t)-\bar{F}(t))(F_{ii}(t_{1})-\bar{F}(t_{1})) }{\sqrt{ \sum_{i} (F_{ii}(t)-\bar{F}(t))^{2} \sum_{i} (F_{ii}(t_{1})-\bar{F}(t_{1}))^{2} }}.
\label{corr}
\end{equation}

\begin{figure}
\begin{center}
\includegraphics[scale=0.65,angle=0]{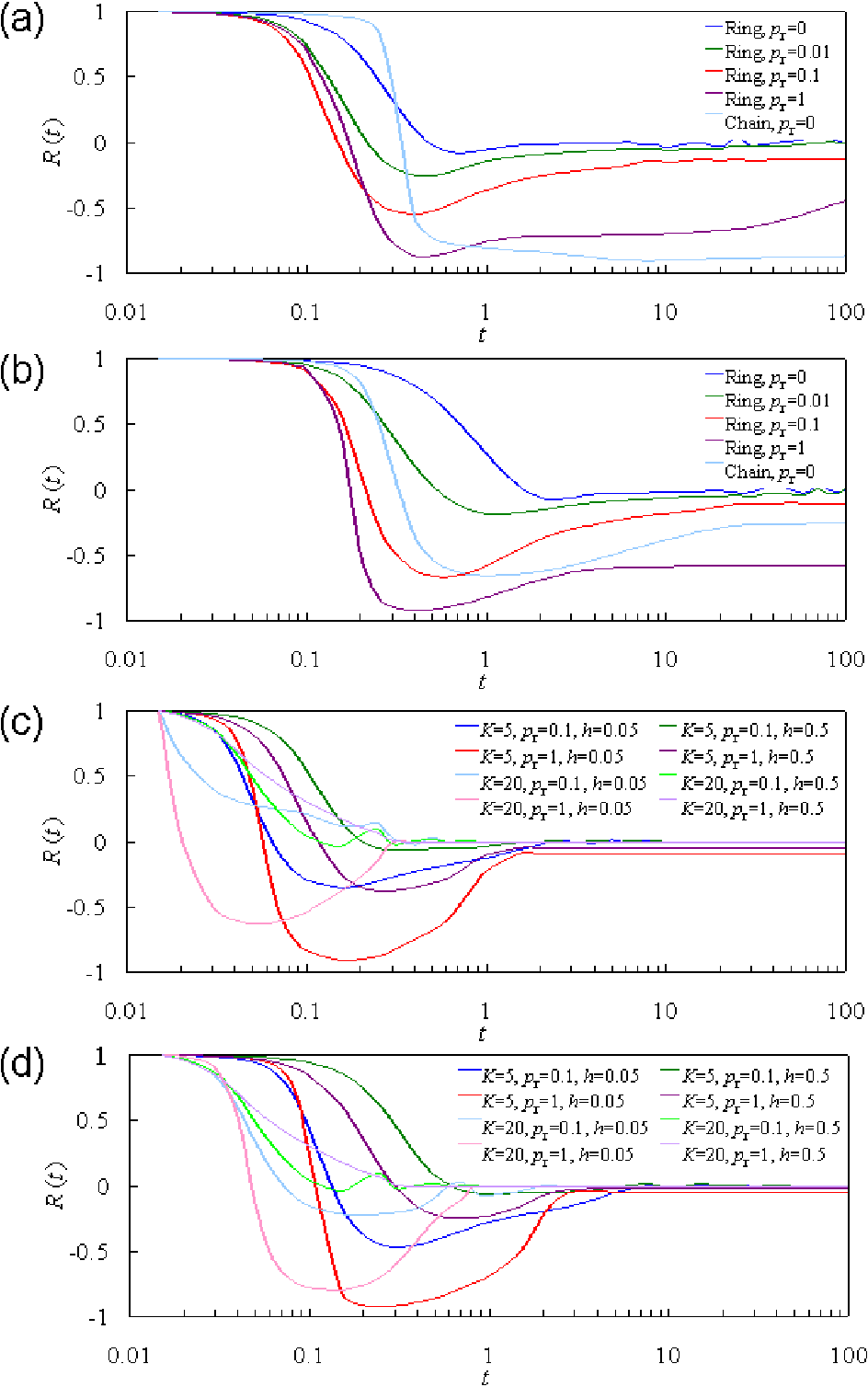}
\end{center}
\caption{$R(t)$ as a function of $t$ for random digraphs of $N=100$. (a) $K=2$ and $h=0.05$ with $\hat{c}=3$ if an arc is present and 0 otherwise. (b) $K=2$ and $h=0.05$ with $\hat{c}=2$ for $j=i$ (self-sustaining profitability), 1 if an arc is present and 0 otherwise. (c) Ring of $K=5$ and 20 with $\hat{c}=3$ if an arc is present and 0 otherwise. (d) Ring of $K=5$ and 20 with $\hat{c}=2$ for $j=i$, 1 if an arc is present and 0 otherwise. For the all graphs, $\hat{d}=1$ and $\hat{y}(0)=10$. The rings with $p_{{\rm r}}=1$ are Erd\"{o}s-R\'{e}nyi random digraphs.}
\label{pre303s}
\end{figure}

Fig. \ref{pre303s} shows $R(t)$. The curves are the average over many different digraphs with the same number of arcs, rewiring, and heterogeneity of parameters. The graph gives the general tendency of the fluctuation as a function of $K$ and $p_{{\rm r}}$ given $h$. A digraph is specified by $K$ and $p_{{\rm r}}$. $K$ is the out-degree of vertices which is the average over the all vertices. Sparse digraphs have a small value of $K$. Some of the arcs are chosen at the rewiring probability of $p_{{\rm r}}$. Either end of the chosen arcs is rewired to a different vertex. The figure shows the curves for rings with $p_{{\rm r}}=0$, 0.01, 0.1, and 1, namely, random digraphs in the Watts-Strogatz model \cite{Wat98}, and chains with $p_{{\rm r}}=0$. The Watts-Strogatz model describes the small-world property of digraphs. The rings with $p_{{\rm r}}=1$ are equivalent to random digraphs in the Erd\"{o}s-R\'{e}nyi model \cite{Erd59}. In the Erd\"{o}s-R\'{e}nyi model, an arc is set between each pair of vertices with equal probability, independently of the other arcs.

In (a), the sign of $R$ turns about sharply after the ignition of profitability. For the rings with small $p_{{\rm r}}$, the initial correlation ($R \sim 1$) is followed by moderate anti-correlation ($R > -0.5$), which disappears as the reciprocal profitability takes over the role to govern $F$ from the profitability and expense. The anti-correlation decreases only very slowly for the rings with $p_{{\rm r}}=1$ where rewiring destroys collectively reciprocal interactions. Strong anti-correlation ($R \sim -1$) appears for the chains where the profitability and expense keep on working for ever. Both the profitability and expense cause anti-correlation, and the anti-correlation is a potential indicator of the transient state. The reason for these is as follows. Initially, a small prosperity means relatively large fluctuation because of the factor $\sqrt{ (d y_{i}(0)+c' y_{i-1}(0) ) }/y_{i}(0)$ in eq.(\ref{fluc_t0}) when $c=0$. After the ignition of the profitability, it mitigates large $F$ for the vertex having small $y(0)$. The expense pressures small $F$ for the vertex having large $y(0)$ into rising. So the anti-correlation appears. At the same time, they tend to compete with each other in forcing the absolute value of the individual $F_{ii}$ to decrease or increase. In (b), anti-correlation disappears gradually as a result of self-sustaining profitability for the chains. This effect is not evident for the rings.

In (c) and (d), if the value of $K$ increases, the value of $p_{{\rm r}}$ increases, or the value of $h$ decreases, the sign of $R$ turns about earlier and subsequent anti-correlation becomes stronger except the case of large $K$ and large $h$ where the anti-correlation nealy vanishes and the correlation converges almost monotonically. The time when the correlation converges also depends on them. If the value of $K$ increases, the value of $p_{{\rm r}}$ decreases, or the value of $h$ increases, the correlation disappears earlier. Such nature of the time evolution of the correlation is demonstrated by the graphs in Figure \ref{pre303s}, which show time constants as a function of $K$, $h$, and $p_{{\rm r}}$. The time when the anti-correlation appears is given by $t_{{\rm a}} = \inf \{ t>0,\ R(t)<0 \}$. The relaxation time for the convergence is defined in this study by $t_{{\rm c}} = \sup \{ t>t_{{\rm a}},\ |R(t)|>0.2 \}$. The convergence of the correlation is slow for digraphs with small $K$ and large $p_{{\rm r}}$. The fluctuation for sparse random digraphs undergoes an upheaval in the salient and long-lived transient state because of the absence of collectively reciprocal interactions. When $p_{{\rm r}}=1$, the probability at which at least one path starting from a vertex forms a ring sub-digraph of length $L$ ($< \log_{K} N$) is given roughly by $p_{{\rm a}}(L) \approx 1-\{(N-2)/(N-1)\}^{K^{L}}$ if $N \gg 1$. When $N=100$ and $K=2$, $p_{{\rm a}}(2) \approx 0.04$. Ring sub-digraphs are missing in the digraph. When $K=5$, $p_{{\rm a}}(2) \approx 0.22$. Ring sub-digraphs appear. When $K=20$, $p_{{\rm a}}(2) \approx 0.98$. Most vertices in the digraph are vertices of ring sub-digraphs. If $K$ is larger, shorter ring sub-digraphs appear more and the convergence of $F$ starts earlier.

\begin{figure}
\begin{center}
\includegraphics[scale=0.45,angle=90]{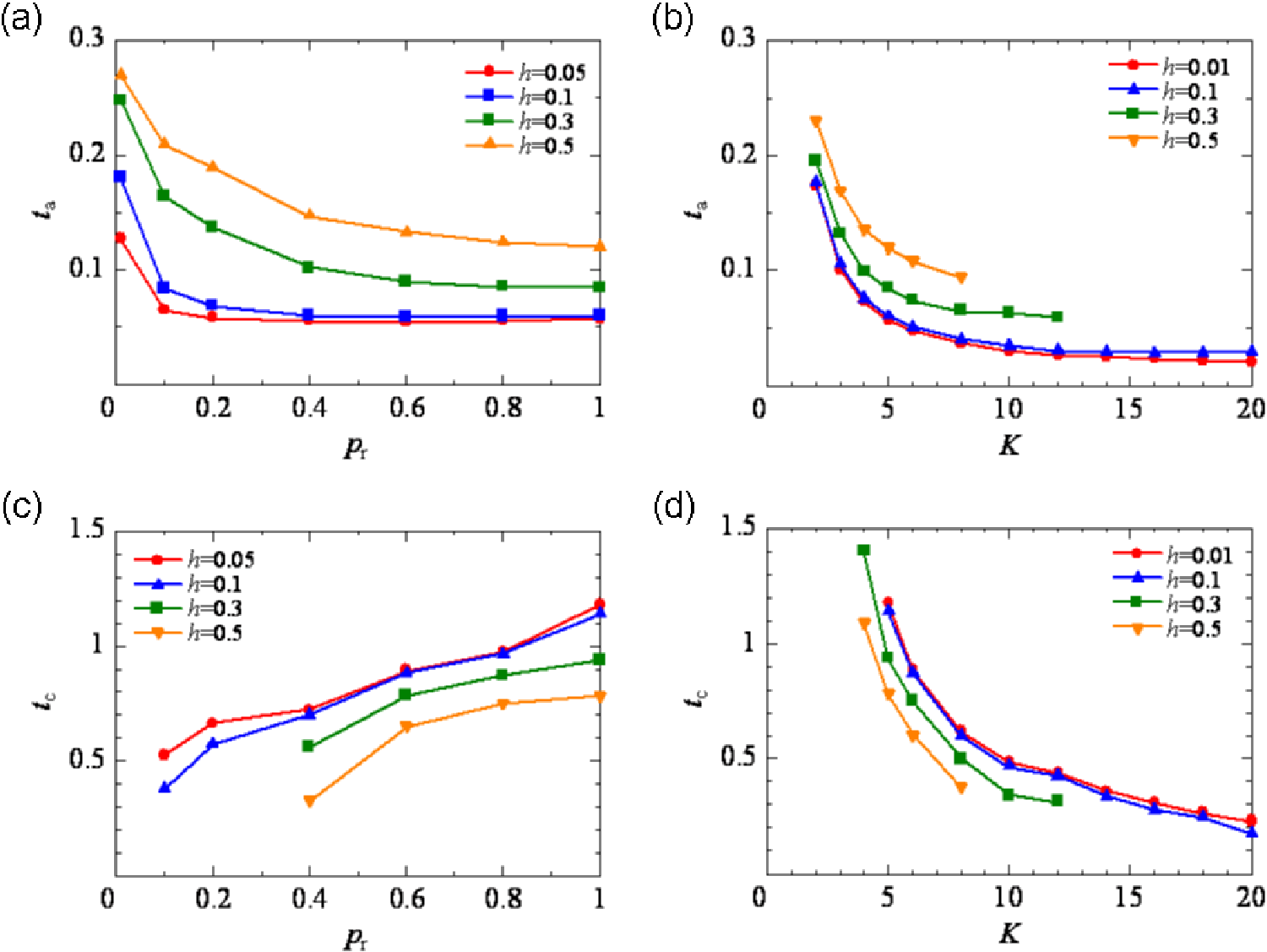}
\end{center}
\caption{Time constants $t_{{\rm a}}$ and $t_{{\rm c}}$ for rings of $N=100$ with $\hat{c}=3$ if an arc is present and 0 otherwise, $\hat{d}=1$, and $\hat{y}(0)=10$. (a) $t_{{\rm a}}$ as a function of $K$ and $h$, $p_{{\rm r}}=1$. (b) $t_{{\rm a}}$ as a function of $p_{{\rm r}}$ and $h$, $K=5$. (c) $t_{{\rm c}}$ as a function of $K$ and $h$, $p_{{\rm r}}=1$. (d) $t_{{\rm c}}$ as a function of $p_{{\rm r}}$ and $h$, $K=5$.}
\label{pre304s}
\end{figure}

\section{Discussion}

The fluctuation does not converge to the limit monotonically. If a firm is a more downstream purchaser (with smaller $i$) along a supply-chain, the expense starts surpassing the profitability there earlier. The supply-chain would not be stable for ever without self-sustaining profitability. This means that there is a finite life time for supply-chains. If the business cluster to which a firm belongs is smaller, the reciprocal profitability is ignited earlier. Just the presence of a business cluster does not predict correctly the risk of impending bankruptcy of individual firms in the transient state. A number of such building blocks are inter-connected with each other to form a general digraph for the entire network economy of firms. The time when the fluctuation converges depends on the sparseness of the general digraph. The swinging changes in $F$ are the consequence of a topology-dependent time trial competition between the profitability and expense. 

No doubt it is a rational strategy to survive in the long run for any firms to occupy a position in a predominant business cluster with reciprocal interactions, say, an intersection of multiple long ring sub-digraphs in the digraph of the entire network economy. But in practice, every time the firms settle accounts, their prosperity must be positive. The selection pressure is brought on the firms nearly continuously if the time interval between their successive settlements is short. Even if a business cluster is present topologically, it may not take effects, nor will it start working soon. The firms must comprehend the time trial competition between the profitability and expense, and determine whether they remain at the current position or bet on the success of rewiring to push back against the immediate selection pressure.

The bankruptcy of a firm ensues directly from an instantaneous fluctuation. It is not within the scope of this study to solve the problem of a first passage time with reflective boundary conditions at $y_{i}=0$, but qualitatively, the bankruptcy of the $i$-th firm is more probable as $F_{ii}$ increases. If $P(\mbox{\boldmath{$y$}},t)$ is approximately a multi-variate normal distribution, the rough estimate of the probability of bankruptcy at $t$ is $p_{{\rm b}}(t) \sim \int_{-\infty}^{0} P_{{\rm M}}(y_{i},t){\rm d}y_{i} = \Phi(-1/F_{ii}(t))$ where $P_{{\rm M}}$ is a marginal distribution of $P$ and $\Phi$ is the cumulative density function of a normal distribution. $F$ in the transient state often exceeds the limit of $F$. The small value of the limit of $F_{ii}$ does not mean the certainty of the survival of the $i$-th firm until then. Some firms are more likely to go bankrupt than expected from the fluctuation in the steady state, and consequently, there is a risk of failing to reach by far the less fluctuating steady state. The risk is more significant in sparse random digraphs. 

%\bibliography{draft0830}.

\end{document}